\documentclass[
reprint,
amsmath,amssymb,
aps,
prb,
]{revtex4-1}

\usepackage{graphicx}
\usepackage{dcolumn}
\usepackage{bm}

\begin{document}

\preprint{APS/123-QED}

\title{Non-Gaussianity in a quasiclassical electronic circuit} 

\author{Takafumi J. Suzuki$^{1,2}$}
\altaffiliation[Present address: ]{
  Department of Physics, Graduate School of Science, The University of Tokyo
}

\author{Hisao Hayakawa$^{2}$}

\affiliation{
  $^1$Department of Physics, Graduate School of Science, The University of Tokyo, 7-3-1 Hongo, Bunkyo-ku, Tokyo 113-0033, Japan \\
  $^2$Yukawa Institute for Theoretical Physics, Kyoto University, Kitashirakawa-oiwake cho, Sakyo-ku, Kyoto 606-8502, Japan
}

\date{\today}

\begin{abstract}

We study the non-Gaussian dynamics of a quasiclassical electronic circuit coupled to a mesoscopic conductor.
Non-Gaussian noise accompanying the nonequilibrium transport through the conductor significantly modifies the stationary probability density function (PDF) of the flux in the dissipative circuit.
We incorporate weak quantum fluctuation of the dissipative LC circuit with a stochastic method, and evaluate the quantum correction of the stationary PDF.
Furthermore, an inverse formula to infer the statistical properties of the non-Gaussian noise from the stationary PDF is derived in the classical-quantum crossover regime.
The quantum correction is indispensable to correctly estimate the microscopic transfer events in the QPC with the quasiclassical inverse formula.

\end{abstract}

\pacs{Valid PACS appear here}
\maketitle


\section{Introduction}

Nonequilibrium fluctuation in mesoscopic conductors has been intensively studied in both classical and quantum regimes
because the fluctuation has essential information on the nonequilibrium transport
\cite{pekola2015towards,Blanter2001}.
Owing to the rapid progress in nanotechnology, it becomes important to address the current distribution beyond the Gaussian one.
In the classical regime, it is experimentally possible to count the number of electrons which pass through a conductor \cite{Fujisawa1634,PhysRevLett.96.076605}.
The achieved histogram characterizes the microscopic transport processes.
The current fluctuation in quantum conductors has also been investigated in terms of the full counting statistics \cite{levitov1996electron,Nazarov2003}.
The current distribution elucidates fundamental aspects of the nonequilibrium properties such as the fluctuation theorem \cite{RevModPhys.81.1665, PhysRevLett.104.080602, PhysRevX.2.011001}.

Non-Gaussianity of nonequilibrium fluctuation has been discussed in various classical systems such as electronic circuits \cite{PhysRevB.80.161203, PhysRevE.90.012115}, diffusive conductors \cite{PhysRevB.66.075334}, chaotic cavities \cite{PhysRevB.66.195318,PhysRevLett.90.206801}, granular particles \cite{0295-5075-102-1-14002,PhysRevLett.110.120601,PhysRevE.94.032910}, nanomagnets \cite{PhysRevLett.114.186601}, particles in dense collides \cite{PhysRevE.94.012610}, and particles with long-range interactions \cite{PhysRevLett.117.030602}.
Recently, a universal mechanism of the non-Gaussianity in such classical systems has been clarified by Kanazawa {\it et al.} \cite{PhysRevLett.114.090601}.
They considered a generic situation where a macroscopic classical particle is coupled to multiple environments, and derived a non-Gaussian Langevin equation even in the thermodynamic limit.
The existence of multiple environments is crucial for the non-Gaussian dynamics of macroscopic systems because it enables different origins of fluctuation and dissipation free of the fluctuation-dissipation theorem.
Otherwise, the noise is reduced to be Gaussian in the macroscopic limit according to the central limit theorem \cite{van1992stochastic}.
They have also demonstrated that the sensitivity of the stochastic particle under the non-Gaussian noise can be utilized to probe the properties of the attaching athermal environment \cite{PhysRevLett.114.090601}.

It is an interesting next step to apply the above discussion to quantum systems.
Unfortunately, the non-Gaussianities in the classical \cite{PhysRevLett.114.090601,PhysRevLett.108.210601} and quantum \cite{levitov1996electron,Nazarov2003} systems have been investigated separately so far,
and the unified understanding on the crossover regime has not yet been established.
This is not just of theoretical interest because recent technology enables us to fabricate electronic nanostructures with high precision.
In particular, on-chip devices are promising systems to investigate the non-Gaussian noise and its quantum effect in a well controlled manner
\cite{PhysRevLett.84.1986,PhysRevLett.96.176601,PhysRevLett.99.206804, hashisaka2008bolometric, ubbelohde2012measurement, jompol2015detecting, PhysRevLett.93.106801, PhysRevLett.93.206601}.
Each element of the electronic circuit is so small that quantum effects play important roles.
Moreover, fluctuation generated in one of the circuit elements is coherently propagated to the rest of the circuit
because the elements embedded on the same chip are strongly coupled with each other.
Thus, systematic studies of quantum transport away from the classical regime is necessary for such small quantum devices.


In this paper, we extend the classical non-Gaussian stochastic equation to the classical-quantum crossover regime.
In order to proceed to concrete discussions,
we consider a dissipative LC circuit inductively coupled with a quantum point contact (QPC).
The LC circuit works in the wide range of scales from the classical \cite{van1992stochastic} to quantum \cite{weiss1999quantum} regimes.
In the classical limit, the LC circuit behaves as a stochastic particle which is subject to both the thermal noise and the non-Gaussian noise generated by the electronic transport through the QPC \cite{PhysRevB.86.075420}.
The system can be also considered as a mesoscopic conductor with a simple and realistic detector circuit to probe the current distribution \cite{lesovik1997detection,PhysRevB.62.R10637,Nazarov2003,PhysRevB.74.115323,PhysRevB.86.075420,PhysRevLett.99.066601,PhysRevB.81.205411}.
Away from the classical limit, the quantum fluctuation which originates from the quantum nature of the dissipative LC circuit becomes relevant to the circuit dynamics.
Hence, the problem should be interpreted as that of a quantum Brownian motion driven by the non-Gaussian noise.
The quantum fluctuation is described by non-Markovian contribution of the Gaussian noise kernel, and is controlled by the quantum-mechanical time scale and a phenomenologically introduced cutoff frequency.
We use a stochastic method to describe the weak quantum fluctuation, and evaluate the quantum correction of the stationary probability density function (PDF) in the quasiclassical regime.

This paper is organized as follows.
In Sec.~\ref{sec: model}, we give a microscopic description of the dissipative LC circuit coupled to the QPC, and simplify the model by assuming the separation of the time scales of the subsystems.
In Sec.~\ref{sec: Langevin equation}, the dynamics of the coupled system is formulated based on a stochastic method.
In Sec.~\ref{sec: Master equation}, the quantum correction of the stationary PDF is determined by solving the Master equation.
In Sec.~\ref{sec: results and discussion}, the effect of the quantum correction and the non-Gaussian noise is numerically evaluated.
We will summarize the results in Sec.~\ref{sec: summary}.
In Appendix \ref{app: transition rate}, we give microscopic expressions of the transition rates for electrons tunneling through the QPC.
In Appendix \ref{app: Derivation of Langevin}, we present the partition function of the stochastic model activated by the non-Gaussian noise.
In Appendix \ref{app:Derivation of Master equation}, we derive the Master equation corresponding to the non-Gaussian Langevin equation.

\section{Model}
\label{sec: model}

\begin{figure}[t]
\centering
\includegraphics[width=0.9\hsize]{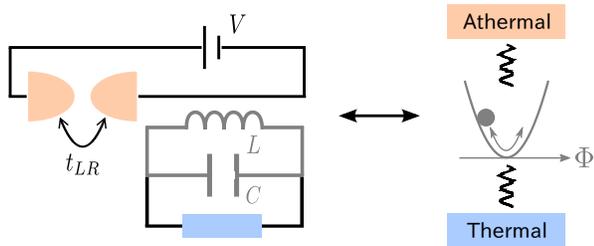}
\caption{
  (Color online)
  Schematic of a dissipative LC circuit inductively coupled to a QPC and its interpretation as a stochastic particle attached to the thermal and athermal environments.
}
\label{fig: system}
\end{figure}

\subsection{Dissipative LC circuit coupled to a QPC}


In this section, we introduce a microscopic model of a dissipative LC circuit coupled with a QPC (Fig.~\ref{fig: system}).
The flux $\Phi$, which is given by the time integral of the voltage $\hat{V}$ across the inductor as $\Phi(t)=\int^{t} dt' \hat{V}(t')$, plays a central role to describe the LC circuit.
A superconducting quantum interference device, which is a typical on-chip device used in various experiments
\cite{deblock2003detection, chiorescu2004coherent, PhysRevLett.93.206601, PhysRevLett.96.127006, Manucharyan113, basset2010emission, houck2012chip, PhysRevLett.108.046802},
behaves as an LC circuit 
when the Josephson coupling energy dominates the charging energy \cite{weiss1999quantum}.
In the superconducting device, the flux $\Phi$ is directly related to the phase difference, which can be observed in terms of the voltage across the junction.
In order to describe the LC circuit attached to a thermal environment, let us consider the Caldeira-Leggett model \cite{caldeira1983path}
\begin{align}
  S_{\rm LC}[\Phi] = \frac{1}{2} \int_{C} dz dz'
  \Phi(z){\mathcal G}^{-1}(z,z')\Phi(z').
  \label{eq:Action of RLC}
\end{align}
The argument $z$ is a combination of the real-time $t$ and the Keldysh index \cite{chou1985equilibrium,kamenev2011field} $\rho=\mp$.
The forward and backward branches of the Keldysh contour $C$ are denoted by $C^{-}$ and $C^{+}$, respectively.
The symbol $\int_{C}$ denotes the integration over the Keldysh contour.
It is convenient to use the action in the rotated Keldysh basis
\begin{align}
  S_{\rm LC}[\Phi^{\rm cl},\Phi^{\rm q}] = \int dt dt'
  \left[
  \Phi^{\rm q}(t) 
  \left({\mathcal G}^{-1}\right)^{\rm r}(t,t')\Phi^{\rm cl}(t')
  \right. \nonumber \\
  \left.
  +\frac{1}{2} \Phi^{\rm q}(t) 
  \left({\mathcal G}^{-1}\right)^{\rm K}(t,t')\Phi^{\rm q}(t')
  \right],
  \label{eq:Action of RLC rotated basis}
\end{align}
with the classical and quantum components of the bosonic field,
$\Phi^{\rm cl} \equiv \left( \Phi^{-}+\Phi^{+}\right) / 2$
and
$\Phi^{\rm q}  \equiv \Phi^{-}-\Phi^{+}$, respectively.
We have introduced the retarded Green's function
\begin{align}
  \left({\mathcal G}^{-1}\right)^{\rm r}(\omega) 
  &= C (\omega^2 -\Omega^2) +i J(\omega),
\end{align}
with the capacitance $C$, the inductance $L$,
the resonant frequency $\Omega\equiv 1/\sqrt{LC}$,
and the spectral function\begin{align}
  J(\omega)= \frac{\sigma\omega}{1+\left(\omega/\omega_{\rm D}\right)^2}.
  \label{eq:Spectral function}
\end{align}
Here, we assume the Ohmic damping with the conductance $\sigma$ and the Drude cutoff frequency $\omega_{\rm D}$.
The Keldysh component of the Green's function in Eq.~(\ref{eq:Action of RLC rotated basis}) is characterized as
\begin{align}
  \left({\mathcal G}^{-1}\right)^{\rm K}(\omega) 
  &= 2i J(\omega) {\rm coth}(\beta\hbar\omega/2)
  \label{eq:inverse Keldysh LC}
\end{align}
when the thermal environment is in equilibrium at inverse temperature $\beta$.
The quantum nature of the dissipative LC circuit is solely incorporated in the fluctuation as a consequence of the linearity of the system \cite{weiss1999quantum}.

The action of the QPC is composed of two parts \cite{kamenev2011field} as $S_{\rm QPC}[\bar{c},c;\Phi]=S_{0}[\bar{c},c]+S_{T}[\bar{c},c;\Phi]$.
The first term $S_{0}$ describes noninteracting electrons in the leads;
\begin{align}
  S_{0}[\bar{c},c]=
  &\sum_{i=L,R}\sum_{\bm k} \int_{C} dz dz' 
  \bar{c}_{i {\bm k}}(z) g^{-1}_{i {\bm k}}(z,z') c_{i {\bm k}}(z'),
  \label{eq: Lead action}
\end{align}
where we have introduced the Grassmann field $\bar{c}_{i {\bm k}}$ ($c_{i {\bm k}}$) for the creation (annihilation) of an electron with the wave vector ${\bm k}$ in the lead $i=L,R$.
We do not consider the spin degrees of freedom for simplicity.
The Green's function in Eq.~(\ref{eq: Lead action}) is defined as
  $g^{-1}_{i{\bm k}}(z,z')= \delta(z-z')\left[  i\hbar\frac{\partial}{\partial z} 
  - \epsilon_{i{\bm k}} \right] $
with the dispersion relation $\epsilon_{i {\bm k}}$ for $i=L,R$.
The bias voltage $V$ is applied to the leads, which are assumed to be in equilibrium at inverse temperature $\beta_{\rm QPC}$.
The second term $S_{\rm T}$ describes the electron hopping which is accompanied with an electronic excitation in the inductively coupled LC circuit \cite{weiss1999quantum};
\begin{align}
  S_{\rm T}[\bar{c},c;\Phi]
  = \frac{1}{N} \sum_{{\bm k},\bm{k'}} \int_{C} dz
  &\left[t_{LR} e^{\frac{ie}{\hbar}\alpha\Phi(z)} \bar{c}_{R{\bm k}}(z) c_{L{\bm k'}}(z) 
    \right. \nonumber \\
    & \hspace{10pt}\left. 
    + {\rm H.c.}
    \right].
\end{align}
Here, ${\rm H.c.}$ denotes the Hermitian conjugate of the first term.
The hopping amplitude between the left and the right leads is denoted by $t_{LR}$.
The coupling constant between the QPC and the LC circuit is given by $\alpha$.

As the QPC action
$S_{\rm QPC}[\bar{c},c;\Phi]=S_{0}[\bar{c},c]+S_{\rm T}[\bar{c},c;\Phi]$
is quadratic in terms of the conduction electrons,
we can exactly perform the Gaussian integration over the fermionic degrees of freedom \cite{nazarov2009quantum}.
The action becomes
\begin{align}
  &S_{\rm QPC}[\Phi] \nonumber \\
  &=i\hbar\int_{C} dz
  \ln\left[ 1-  t_{LR}^2 G_{L} V[\Phi] G_{R} V^{\dagger}[\Phi]\right](z,z),
  \label{eq:effective action for QPC}
\end{align}
where we have introduced the vertex operator
\begin{align}
  V[\Phi]\equiv \exp\left( \frac{ie\alpha}{\hbar} \Phi \right).
\end{align}
The integration on the Keldysh contour is abbreviated as
$(AB)(z,z')\equiv\int_{C} dz'' A(z,z'')B(z'',z')$.
The retarded and the Keldysh components of the Green's function $G_{i}(z,z')\equiv\frac{1}{N}\sum_{{\bm k}}g_{i{\bm k}}(z,z')$ are, respectively, given by 
\begin{align}
  \label{eq: QPC retarded}
  &G^{\rm r}_{i}(\omega) =  -i \pi \rho_{i} , \\
  \label{eq: QPC Keldysh}
  &G^{\rm K}_{i}(\omega) =  2\pi i \rho_{i} (2f_{i}(\omega)-1),
\end{align}
where $f_{i}(\omega)=1/(\exp\left[\beta_{\rm QPC}(\omega-\mu_{i})\right]+1)$ is the Fermi-Dirac distribution of the lead $i=L$, $R$ with the chemical potentials $\mu_{L}=eV$ and $\mu_{R}=0$.
The energy dependence of the density of states $\rho_{i}$ ($i=L$, $R$) is neglected in Eqs.~(\ref{eq: QPC retarded}) and (\ref{eq: QPC Keldysh}) for simplicity.


\subsection{Quasistationary approximation}
\label{subsec: quasistationary}

In this paper, we assume that a typical time scale of the QPC is much shorter than that of the LC circuit.
The correlation time of the current fluctuation in the QPC is governed by the bias voltage $V$ and the inverse temperature $\beta_{\rm QPC}$,
while the dynamics of the LC circuit is characterized by the resonant frequency $\Omega$ and the relaxation time $C/\sigma$.
If the dynamics of the subsystems is well separated, we are allowed to consider an intermediate time scale $\Delta t$ satisfying
\begin{align}
  \min \left( \frac{\hbar}{eV} , \hbar\beta_{\rm QPC} \right)
  \ll \Delta t
  \ll \min \left( \frac{C}{\sigma} , \frac{1}{\Omega}
  \right).
  \label{eq: quasistationary approximation}
\end{align}
Within the quasistationary approximation \cite{PhysRevB.86.075420,PhysRevLett.93.106801,PhysRevLett.90.206801,ANDP:ANDP200710259}, we can regard the vertex operator $V[\Phi]$ in the QPC action (\ref{eq:effective action for QPC}) as a constant because the dynamics in the LC circuit is almost frozen during $\Delta t$.
At the same time, $\Delta t$ is so long for the QPC dynamics that the current fluctuation in the QPC is characterized by the stationary distribution.
This argument can be further substantiated by the path-integral representation \cite{PhysRevB.86.075420,ANDP:ANDP200710259} with the discrete time stride $\Delta t$.
Thus, we find that the QPC action has the same structure as the cumulant generating functional in the full counting statistics \cite{levitov1996electron} as
\begin{align}
  &S_{\rm QPC}[\Phi^{\rm q}]  \nonumber \\
  \label{eq: Levitov-Lesovik}
  &=i\hbar \int dt \int \frac{d\omega}{2\pi}
  \ln \left[ 
    1 + T_{LR}
    \left[
    f_L(\omega) (1-f_R(\omega)) 
    \right.
    \right.
    \nonumber \\
    & \hspace{5pt}
    \left.
    \left.
    \times (e^{\frac{ie\alpha}{\hbar}\Phi^{\rm q}(t)}-1)
    +f_R(\omega) (1-f_L(\omega)) 
    (e^{-\frac{ie\alpha}{\hbar}\Phi^{\rm q}(t)}-1)
    \right]
    \right] \nonumber \\
  &\equiv-i \sum^{\infty}_{n=-\infty}\int dt 
    W_{n} (e^{\frac{ie\alpha}{\hbar}n\Phi^{\rm q}(t)}-1),
\end{align}
with the transmission coefficient $T_{LR}\equiv 4\pi^2 t^2_{LR}\rho_L\rho_R$.
The last line of Eq.~(\ref{eq: Levitov-Lesovik}) defines the transition rate $W_{n}$ as a coefficient of $(e^{\frac{ie\alpha}{\hbar}n\Phi^{\rm q}(t)}-1)$ (see Appendix \ref{app: transition rate} for the explicit expressions of the lowest-order terms).

With the action of the LC circuit (\ref{eq:Action of RLC rotated basis}) and the QPC (\ref{eq: Levitov-Lesovik}), we obtain the partition function
\begin{align}
  Z = \int {\cal D}\Phi^{\rm cl} {\cal D}\Phi^{\rm q} \exp{ \left[\frac{i}{\hbar}\left(S_{\rm LC}[\Phi^{\rm cl},\Phi^{\rm q}]+S_{\rm QPC}[\Phi^{\rm q}]\right)\right]}.
  \label{eq: partition function}
\end{align}
The important point is that the QPC action no longer depends on $\Phi^{\rm cl}$ as a consequence of the time-scale separation.
This indicates that
the instantaneous current through the QPC perturbs the flux in the LC circuit, while the current distribution is not influenced by the state of the LC circuit.

\section{Stochastic formulation}
\label{sec: Langevin equation}

In this section, we develop a stochastic method for the QPC-LC coupled system by relating the partition function (\ref{eq: partition function}) with characteristic functionals of stochastic processes within a quasiclassical approximation.
It is found that the electron transport in the QPC perturbs the LC circuit with a compound Poisson noise, which is a typical non-Gaussian noise.
The thermal and quantum fluctuations of the dissipative circuit can be incorporated in the noise kernel.
Asymptotic-scaling analysis of the non-Gaussian stochastic equation is performed as a quantum extension of Ref.~\onlinecite{PhysRevLett.114.090601}.
We reduce the non-Markovian dynamics of the quantum LC circuit to multivariate Markovian processes by decomposing the quantum fluctuation into exponentially-correlated auxiliary variables.

\subsection{Characteristic functionals}

The statistical property of the non-Gaussian noise produced by the QPC is completely described by a characteristic functional.
In order to recover the thermal fluctuation in the classical limit \cite{kamenev2011field}, we scale the fields as
$\Phi^{\rm q}(t) \equiv (\hbar /e\gamma) \varphi^{\rm q}(s)$
and
$\Phi^{\rm cl}(t) \equiv (e/C \Omega)\varphi^{\rm cl}(s)$
with the dimensionless time
$s \equiv \Omega t $
and the dimensionless friction coefficient
$\gamma\equiv \sigma/C\Omega$.
The contribution from the QPC in the partition function is given by
\begin{align}
  &\chi_{\rm NG} [\varphi^{\rm q}] \equiv  \exp \left[ \lambda_{\rm P}
    \int ds \left[
      \int dy w(y) \left( e^{iy \varphi^{\rm q}(s) }- 1 \right) \right]
    \right],
  \label{eq: chi_NG}
\end{align}
where we have introduced the rate parameter $\lambda_{\rm P}\equiv \sum_{n}w_{n}$
and the jump size distribution
\begin{align}
  w(y)\equiv \frac{1}{\lambda_{\rm P}} \sum^{\infty}_{n=-\infty}
  w_{n} \delta\left(y- \frac{n\alpha}{\gamma}\right),
  \label{eq: PD of shot noise}
\end{align}
with the dimensionless transition rate
$w_{n}\equiv W_{n}/\hbar\Omega$ ($n\in \mathbb{Z}$).
The functional $\chi_{\rm NG}$ is identical to the characteristic fuctional of compound Poisson processes \cite{hanggi1989path}.
The physical processes behind the generation of the non-Gaussian noise are explained as follows.
The transport processes in the QPC are composed of the independent microscopic events where several electrons are transferred in a short time interval.
The probability for $n$ electrons transferred from the left reservoir to the right one is proportional to the transition rate $w_{n}$.
According to Ampere's law, the instantaneous current though the QPC produces a magnetic flux whose amplitudes are proportional to the current value.
Due to the granularity of electrons, the possible amplitudes of the non-Gaussian noise induced in the LC circuit are integer multiples of the unit amplitude $\alpha/\gamma$.
The statistical properties of the non-Gaussian noise are characterized as the weighted summation of these microscopic events.

The Gaussian fluctuation which originates from the linear coupling between the circuit and the thermal environment is described by the quadratic term \cite{kamenev2011field} in $\varphi^{\rm q}$ in the LC action (\ref{eq:Action of RLC rotated basis}).
This contribution is identical to the Gaussian characteristic functional
\begin{align}
  &\chi_{\rm G} [\varphi^{\rm q}] \equiv \exp \left[ - \int ds ds' \varphi^{\rm q}(s) \nu(s-s') \varphi^{\rm q}(s') \right],
  \label{eq: chi_G}
\end{align}
with the noise kernel \cite{weiss1999quantum}
\begin{align}
    \nu(s) 
    &\equiv \frac{\hbar C^2}{2e^2\sigma^2}
    \int ^{\infty}_{-\infty}\frac{d\omega}{2\pi} J(\omega) \coth\left( \frac{\beta\hbar\omega}{2}\right)e^{-is \omega / \Omega}.
    \label{eq: def. noise kernel}
\end{align}
In the case of the Ohmic spectral function with the Drude regularization (\ref{eq:Spectral function}), the noise kernel becomes
\begin{align}
   \nu(s) =
  \frac{T}{\gamma}
  &\left[  
    \frac{\omega_{\rm c}}{2}e^{-|s|\omega_{\rm c}}
    +\sum^{\infty}_{k=1}\frac{1}{(k /\tau\omega_{\rm c})^2-1} 
    \right.
    \nonumber \\
    & \hspace{30pt} \times
    \left.
    \left(
    \omega_{\rm c}e^{-|s|\omega_{\rm c}}
    -\frac{k}{\tau}e^{-k|s|/\tau}
    \right)
    \right],
  \label{eq:Noise kernel}
\end{align}
where $T\equiv C/e^2\beta$, $\omega_{\rm c}\equiv \omega_{\rm D}/\Omega$, and $\tau\equiv\beta\hbar\Omega/2\pi$
are the dimensionless temperature, the cutoff frequency, and the quantum-mechanical scale, respectively.
In the following discussions, the cutoff frequency $\omega_{\rm D}$ is assumed to be much larger than the resonant frequency $\Omega$, i.e. $\omega_{\rm c} \gg 1$.
In this regime, the first term on the right-hand side (RHS) of Eq.~(\ref{eq:Noise kernel}) is reduced to the thermal noise $(T/\gamma) \delta(s)$.
The second term on the RHS of Eq.~(\ref{eq:Noise kernel}) describes the non-Markovian quantum fluctuation.
In the classical limit ($\tau=0$), the second term vanishes and the noise kernel is solely determined by the thermal noise.
As will be shown later, the existence of the cutoff and the exponentially decaying form of the second term is important to discuss the quantum correction.

In the subsequent discussions, we focus on the overdamped case ($\gamma > 1$).
The partition function (\ref{eq: partition function}) is written in terms of the two characteristic functionals Eqs.~(\ref{eq: chi_NG}) and (\ref{eq: chi_G}) as
\begin{align}
  Z=&\int {\cal D}\varphi^{\rm cl}{\cal D}\varphi^{\rm q}
  \chi_{\rm G}[\varphi^{\rm q}]\chi_{\rm NG}[\varphi^{\rm q}] 
  \nonumber \\
  & \exp \left[ i\int ds 
    \varphi^{\rm q}(s) 
    \left( 
    - \frac{\partial}{\partial s} - \frac{1}{\gamma}
    \right)
    \varphi^{\rm cl} (s) \right].
  \label{eq: partition function with CGFs}
\end{align}
In the derivation the equation of motion of the LC circuit,
we have considered that the friction kernel becomes memoryless under assumption of the large cutoff frequency $\omega_{\rm c} \gg 1$.



\subsection{Asymptotic scaling in the classical limit}
\label{subsec: classical limit}

In the classical limit, the partition function (\ref{eq: partition function with CGFs}) leads to the over-damped Langevin equation with the zero-mean thermal noise $\eta_{0}$ (Appendix \ref{app: Derivation of Langevin}) and the compound Poisson noise \cite{hanggi1989path};
\begin{align}
  &\frac{\partial\varphi^{\rm cl}(s)}{\partial s} = 
  - \frac{1}{\gamma} \varphi^{\rm cl}(s)
  + \sqrt{\frac{T}{\gamma}} \eta_{0}(s) 
  +  \sum_{i} y_{i}\delta(s-s_{i}), \\
  &\langle \eta_{0}(s)\eta_{0}(s')\rangle = 2\delta(s-s').
  \label{eq: thermal noise}
\end{align}
The statistical properties of the non-Gaussian noise are characterized by the non-Gaussian characteristic functional $\chi_{\rm NG}$ (\ref{eq: chi_NG}).
A set of the arrival times $\{s_{i}\}$ obeys the Poisson process with a rate parameter $\lambda_{\rm P}$.
Each amplitude of the noise $y_{i}$ is independent and identically distributed random variable with the jump size distribution $w(y)$.
In our model, each amplitude takes $n\alpha/\gamma$ with probability $w_{n}/\lambda_{\rm P}$.

The previous work \cite{PhysRevLett.114.090601} pointed out that the non-Gaussianity survives even in the thermodynamic limit under the following three assumptions:
(i) the amplitude of the non-Gaussian noise is small;
(ii) the friction coefficient $\gamma$ is independent of the non-Gaussian-noise amplitude;
(iii) the thermal noise is smaller than or of the same order as the non-Gaussian noise.
In our case, the condition (i) is satisfied if there exists an integer $n_{c}\in \mathcal{N}$ such that (a) $w_{|n|>n_{c}} / \lambda_{\rm P}\ll 1$ and (b) $n_{c} \alpha / \gamma \ll 1$.
The condition (a) holds at weak-tunneling regime ($T_{LR} \ll 1$) because the transport processes where many electrons are transmitted instantaneously are rare events.
The condition (b) requires the characteristic amplitude $\epsilon \equiv \alpha/\gamma$ of the non-Gaussian noise to be small for dominant transport processes.
The conditions (ii) and (iii) are straightforwardly translated in our system.
The condition (ii) is satisfied because the friction coefficient $\gamma$ is irrelevant to the athermal environment.
Besides, the condition (iii) holds when the scaled temperature ${\mathcal T} \equiv  T/ \epsilon^2$ is independent of $\epsilon$.

Following the previous work, we use the rescaled variables
  $\phi\equiv\varphi^{\rm cl}/\epsilon$
and
  ${\mathcal Y}\equiv y/\epsilon$
to obtain the asymptotic Langevin equation
\begin{align}
  \label{eq: classical NG Langevin}
  &\frac{\partial\phi(s)}{\partial s} = 
  - \frac{1}{\gamma} \phi(s)
  + \sqrt{\frac{{\mathcal T}}{\gamma}} \eta_{0}(s) 
  + \eta_{\rm P}(s),
\end{align}
with the thermal noise $\eta_{0}$ and the non-Gaussian noise
\begin{align}
  \eta_{\rm P}(s)=\sum_{i} {\mathcal Y}_{i}\delta(s-s_{i}),
  \label{eq: rescaled shot noise}
\end{align}
characterized by the rate parameter $\lambda_{\rm P}$ and the rescaled jump size distribution
\begin{align}
  {\mathcal W}({\mathcal Y}) 
  & \equiv \epsilon w(y) \nonumber \\
  &=  \frac{1}{\lambda_{\rm P}} \sum^{\infty}_{n=-\infty}
  w_{n} \delta({\mathcal Y}-n).
   \label{eq: scaled PD of shot noise}
\end{align}
The existence of the non-Gaussian Langevin equation (\ref{eq: classical NG Langevin}) and its stationary solution have already been established by the previous work \cite{PhysRevLett.114.090601}.
In the path-integral formalism, the stochastic differential equation (\ref{eq: classical NG Langevin}) corresponds to the non-Gaussian action
\begin{align}
  \label{eq: continuous action}
  S=\int ds &\left[ -\phi^{\rm q}(s) 
    \left[   
      \frac{\partial}{\partial s}+\frac{1}{\gamma}  
      \right] \phi(s)
    +i\frac{\mathcal{T}}{\gamma} \left( \phi^{\rm q}(s) \right)^2 \right. \nonumber \\
    &
    -i  \lambda_{\rm P}
    \int d{\mathcal Y} \mathcal{W}(\mathcal{Y}) \left. \left( e^{i \mathcal{Y}\phi^{\rm q}(s)} - 1 \right)  \right],
\end{align}
with $\phi^{\rm q} \equiv \varphi^{\rm q}/\epsilon$.


\subsection{Quantum fluctuation}

The quantum nature of the dissipative LC circuit becomes relevant at low temperatures where the correlation time $\beta\hbar$ of the thermal environment is comparable to the inverse of the resonant frequency $1/\Omega$.
In the following, we consider the regime where the quantum-mechanical correlation time $\beta\hbar/2\pi$ is much larger than the QPC time scales $\hbar/eV$ and $\hbar\beta_{\rm QPC}$ so that the quasistationary approximation (\ref{eq: quasistationary approximation}) is applicable
\footnote{
If the time scale of the LC circuit is comparable to that of the QPC, we can not use the approximation and need to solve the quantum dynamics of the coupled system.}
.
It is still not necessary to consider the memory effect of the friction kernel as long as the cutoff frequency is sufficiently large ($\omega_{\rm c} \gg 1$).

The quantum effect is incorporated via the second term on the RHS of Eq.~(\ref{eq:Noise kernel}), whose $k$-th term is
\begin{align}
   \frac{\mathcal T}{\gamma}\frac{1}{(k/\tau\omega_{\rm c})^2-1}
    \left(
    \omega_{\rm c}e^{-\omega_{\rm c}|s|}
    -\frac{k}{\tau}e^{-k|s|/\tau}
    \right).
    \label{eq: kth noise kernel}
\end{align}
By applying the Hubbard-Stratonovich transformation, we can decompose the quantum correction into mutually independent auxiliary variables $\eta^{+}_{k}$ and $\eta^{-}_{k}$
which have exponentially decaying correlations.
In other words, if we assume that the auxiliary variables $\eta^{\mu=\pm}_{k}$ obey the Ornstein-Uhlenbeck processes
\begin{align}
  \label{eq: SDE of eta}
  &\frac{\partial \eta^{\mu}_{k}(s)}{\partial s} =
  \frac{-\eta^{\mu}_{k}(s)+\xi^{\mu}_{k}(s)}{\tau^{\mu}_{k}},\\
  \label{eq: statistics of eta}
  &\langle \xi^{\mu}_{k}(s)\xi^{\mu'}_{k'}(s')\rangle
  = 2\delta(s-s') \delta_{kk'}\delta_{\mu\mu'},
\end{align}
with $\tau^{+}_{k}\equiv 1/\omega_{\rm c}$, $\tau^{-}_{k}\equiv \tau/k$, and the Kronecker delta $\delta_{ij}$,
their solutions have the same statistical properties as the original fluctuation as
\begin{align}
  \langle \eta^{\mu}_{k}(s) \eta^{\mu'}_{k'}(s')\rangle
  &= \frac{1}{\tau^{\mu}_{k}}\exp\left({-|s-s'|/\tau^{\mu}_{k}}\right)
  \delta_{kk'}\delta_{\mu\mu'}.
  \label{eq: persistent noise}
\end{align}
Hence, it is possible to consider that the flux $\phi$ is driven by the auxiliary variables $\eta^{\mu}_{k}$.
It is interesting that the persistent noise (\ref{eq: persistent noise}) plays an important role in active matter \cite{PhysRevLett.117.038103} as well.
The overdamped Langevin equation with the quantum correction is written as
\begin{align}
  \label{eq: Langevin with shot and quantum noise}
  \frac{\partial\phi(s)}{\partial s} 
  =& 
  - \frac{1}{\gamma} \phi(s)
  + \sqrt{ \frac{{\mathcal T}}{\gamma}} \eta_{0}(s) 
  + \eta_{\rm P}(s) \nonumber \\
  &+ \sum^{\infty}_{k=1} \sqrt{ \frac{{\mathcal T}_{k}}{\gamma}}
  \left[ \eta^{+}_{k}(s)+i\eta^{-}_{k}(s)\right],
\end{align}
with ${\mathcal T}_{k}={\mathcal T}/((k/\tau\omega_{\rm c})^2-1)$.
We note that the $\eta^{-}_{k}$-term is purely imaginary because the second term of Eq.~(\ref{eq: kth noise kernel}) is negative.
To summarize, the quantum dynamics of the QPC-LC coupled system is mapped to the linearly-coupled Langevin equations
 with the auxiliary variables.

\section{Stationary PDF}
\label{sec: Master equation}

In this section, we analyze the stochastic equations Eqs.~(\ref{eq: thermal noise}), (\ref{eq: rescaled shot noise}), (\ref{eq: persistent noise}) and (\ref{eq: Langevin with shot and quantum noise}) by switching to the Master equation approach (Appendix \ref{app:Derivation of Master equation}).
We denote the probability density function (PDF) of the quasiclassical LC circuit
\footnote{
The probabilistic interpretation of the PDF fails in the quantum regime.
In this case, we need the quantum mechanical treatment of the density matrix of the circuit \cite{Nazarov2003}.
}
by ${\mathcal P}(\phi,\bm{\eta},s)$ with the flux $\phi$, a set of the auxiliary variables ${\bm \eta}$, and the rescaled time $s$.
The corresponding Master equation is 
\begin{widetext}
  \begin{align}
    \frac{\partial {\mathcal P}(\phi,\bm{\eta},s)}{\partial s}
    =&
    \frac{1}{\gamma}\frac{\partial}{\partial \phi}\left[ \phi {\mathcal P}(\phi,\bm{\eta},s)\right]
    +\frac{{\mathcal T}}{\gamma} 
    \frac{\partial^2 {\mathcal P}(\phi,\bm{\eta},s)}{\partial \phi^2}   
    +\lambda_{\rm P}\int d{\mathcal Y} {\mathcal W}({\mathcal Y})
    \left( e^{-{\mathcal Y}\frac{\partial}{\partial \phi}} -1\right) {\mathcal P}(\phi,\bm{\eta},s)
    \nonumber \\
    &
    +\sum^{\infty}_{k=1}\sum_{\mu=\pm} \left(
    \sqrt{\frac{\mu{\mathcal T}_{k}}{\gamma}} \eta^{\mu}_k
    \frac{\partial}{\partial \phi} 
    +\frac{1}{\tau^{\mu}_{k}}
    \frac{\partial}{\partial \eta^{\mu}_{k}}\left[
      \eta^{\mu}_{k} +\frac{1}{\tau^{\mu}_{k}}\frac{\partial}{\partial \eta^{\mu}_{k}}
      \right]\right)
         {\mathcal P}(\phi,\bm{\eta},s).
         \label{eq: master eq. with OU-noise}
  \end{align}  
\end{widetext}
The first line on the RHS of Eq.~(\ref{eq: master eq. with OU-noise}) is identical to the Master equation in the classical regime.
The second line describes the coupling between the flux $\phi$ and the auxiliary variables $\eta^{\mu}_{k}$ and the diffusion of the auxiliary variables.

In the quasiclassical regime where the time constants and the coupling constants are sufficiently small, i.e. $\tau^{\mu}_{k} \ll 1$ and ${\mathcal T}_{k} \ll 1$ for an arbitrary natural number $k \in {\mathcal N}$ and $\mu=\pm$,
the Master equation (\ref{eq: master eq. with OU-noise}) is reduced to a single-variate Fokker-Planck equation \cite{kamenev2011field}.
The former condition is compatible with $1 \ll \omega_{\rm c}$ and $\tau \ll 1$.
The latter one further requires the cutoff frequency to filter the quantum-mechanical energy scale ($\tau\omega_{\rm c}< 1$)
so that the quantum effect remains weak (${\mathcal T}_{k} \ll 1$).
\begin{widetext}
Within this parameter regime, we obtain a perturbative solution around the classical limit as
\begin{align}
  {\mathcal P}(\phi,\bm{\eta},s) 
  \simeq 
  &\exp\left[-\sum^{\infty}_{k=1} \sum_{\mu=\pm}
    \frac{\tau^{\mu}_{k}\left(\eta^{\mu}_{k}\right)^2}{2}
    \right]
  \left[ {\mathcal P}(\phi,s)
    +\sum^{\infty}_{k=1}  \sum_{\mu=\pm}
    \eta^{\mu}_{k} N^{\mu}_{k}(\phi,s)
    \right].
  \end{align}
The coupled Master equations for ${\mathcal P}(\phi,s)$ and $N^{\mu}_{k}(\phi,s)$ are given by
\begin{align}
    \frac{\partial {\mathcal P}(\phi,s)}{\partial s}
    =& \left[
    \frac{1}{\gamma} \left( 1+ \phi \frac{\partial}{\partial \phi} \right) 
    +\frac{{\mathcal T}}{\gamma} 
    \frac{\partial^2}{\partial \phi^2} 
    +\lambda_{\rm P}\int d{\mathcal Y} {\mathcal W}({\mathcal Y})
    \left( e^{-{\mathcal Y}\frac{\partial}{\partial \phi}} -1\right) 
    \right]{\mathcal P}(\phi,s)
    + \sum^{\infty}_{k=1}\sum_{\mu=\pm} 
    \frac{1}{\tau^{\mu}_{k}} \sqrt{ \frac{\mu{\mathcal T}_{k}}{\gamma}}
    \frac{\partial N^{\mu}_{k}(\phi,s)}{\partial \phi} 
     , \\
    \tau^{\mu}_{k}\frac{\partial  N^{\mu}_{k}(\phi,s)}{\partial s}
    =&\left[
      \frac{1}{\gamma}\left( 1+ \phi \frac{\partial}{\partial \phi} \right)
    +\frac{{\mathcal T}}{\gamma} 
    \frac{\partial^2}{\partial \phi^2}
    +\lambda_{\rm P}\int d{\mathcal Y} {\mathcal W}({\mathcal Y})
    \left( e^{-{\mathcal Y}\frac{\partial}{\partial \phi}} -1\right)
    -\frac{1}{\tau^{\mu}_{k}} 
    \right] N^{\mu}_{k}(\phi,s)
    +  \sqrt{\frac{\mu{\mathcal T}_{k}}{\gamma}}
    \frac{\partial {\mathcal P}(\phi,s)}{\partial \phi} 
    ,
\end{align}
respectively.
The equation for the steady-state PDF ${\mathcal P}_{\rm SS}(\phi)$ is given up to the first order in $\tau^{\mu}_{k}$ as
\begin{align}
  0 =
  &\left[
    \left( 1+\phi\frac{\partial}{\partial\phi}\right)
  +{\mathcal T}
  \frac{\partial^2}{\partial \phi^2}
  +  \gamma \lambda_{\rm P}
  \int d{\mathcal Y} {\mathcal W}({\mathcal Y})
  \left(e^{-{\mathcal Y}\frac{\partial}{\partial \phi}}-1\right)
  - \left[ 1-\sum_{k}\frac{{\mathcal T}_{k}\tau_{k}}{\gamma}
    \frac{\partial^2}{\partial \phi^2}\right]^{-1} 
  \sum_{k}{\mathcal T}_{k} \left[ 1-\frac{\tau_{k}}{\gamma}\right]
  \frac{\partial^2}{\partial \phi^2} 
  \right]{\mathcal P}_{\rm SS}(\phi),
  \label{eq: 1st order PDF}
\end{align}
\end{widetext}
with $\tau_{k} \equiv 1/\omega_{\rm c}-\tau/k$.
With the Fourier transform of the steady-state PDF $\tilde{{\mathcal P}}_{\rm SS}(\lambda)$
and the cumulant generating function $\tilde{{\mathcal F}}_{\rm SS}(\lambda)\equiv \ln{\tilde{{\mathcal P}}_{\rm SS}(\lambda)}$,
the stationary solution of Eq.~(\ref{eq: 1st order PDF}) can be written as
\begin{align}
  \tilde{{\mathcal F}}_{\rm SS}(\lambda)
  = \ln{\tilde{\mathcal{K}}(\lambda)} 
  + \tilde{{\mathcal F}}^{\rm cl}_{\rm SS}(\lambda).
  \label{eq: F_ss}
\end{align}
Here, the second term represents the solution in the classical limit \cite{PhysRevLett.114.090601} as
\begin{align}
  \tilde{{\mathcal F}}^{\rm cl}_{\rm SS}(\lambda) 
  = - \frac{{\mathcal T} \lambda^2}{2}
  +\gamma \lambda_{\rm P} \int d{\mathcal Y} {\mathcal W}({\mathcal Y})
  \int^{\lambda}_{0} d\lambda'\frac{e^{i{\mathcal Y} \lambda'} - 1 }{\lambda'}.
  \label{eq: F_classical}
\end{align}
The first term on the RHS of Eq.~(\ref{eq: F_ss}) contains the quantum-correction kernel
\begin{align}
  &\tilde{\mathcal{K}}(\lambda)\equiv \left( \frac{1}{1+\theta^2\lambda^2}\right)^{k},
  \label{eq: quantum kernel FT}
\end{align}
with the parameters
\begin{align}
  \label{eq: scale parameter}
  &\theta\equiv\left(\frac{{\mathcal T}\tau}{\gamma} \left[ \psi\left(1+\tau\omega_{\rm c}\right)-\psi(1)\right] \right)^{1/2}, \\
  \label{eq: shape parameter}
  &k\equiv
  \frac{1}{2}\left[\frac{{\mathcal T} }{2\theta^2}
    \left(
    1-\frac{\pi\tau\omega_{\rm c}}
    { \tan\left(\pi\tau\omega_{\rm c}\right)}
    \right)
    -\frac{1}{\gamma}
    \right].
\end{align}
Here, we have introduced the digamma function $\psi(x) \equiv \frac{d}{dx}\ln \Gamma(x)$ with the gamma function $\Gamma(x)\equiv \int^{\infty}_{0} dt~t^{x-1}e^{-t}$.
The factor of $\tilde{\mathcal{K}}(\lambda)$ is determined by the normalization of the PDF, i.e. $\tilde{{\mathcal P}}_{\rm SS}(0) = 1$.
The quantum-correction kernel (\ref{eq: quantum kernel FT}) is identical to the characteristic functional of the difference of the two independent random variables obeying the identical Gamma distribution with the scale parameter $\theta$ and the shape parameter $k$.
The parameters $\theta$ and $k$ are associated with the amplitude and the arrival rate of the quantum fluctuation, respectively.
The quantum correction does not modulate the mean but the fluctuation of the flux $\phi$.
The quantum correction vanishes in the classical limit, i.e. $\ln{\tilde{\mathcal{K}}(\lambda)}=0$ for $\tau=0$.

The kernel $\tilde{\mathcal{K}}(\lambda)$ can be Fourier transformed for $k>0$ as
\begin{align}
  \mathcal{K}(\phi)&=\frac{1}{\sqrt{\pi}\Gamma(k)\theta}
  \left( \frac{|\phi|}{2\theta} \right)^{k-1/2}
  K_{k-1/2}\left(|\phi| / \theta \right),
  \label{eq: quantum kernel}
\end{align}
with the modified Bessel function of the second kind $K_{\nu}(z)$.
The stationary PDF with the quantum correction $\mathcal{K}(\phi)$ is related with the classical solution as 
\begin{align}
  {\mathcal P}_{\rm SS}(\phi)  = \int d\phi' \mathcal{K}(\phi-\phi'){\mathcal P}^{\rm cl}_{\rm SS}(\phi').
 \label{eq: PDF with quantum kernel}
\end{align}
Thus, the kernel $\mathcal{K}(\phi)$ characterizes the quantum correction.
This is one of the main results of this paper.

As was discussed in the previous work \cite{PhysRevLett.114.090601},
the stationary PDF of the classical particle can be utilized to probe the non-Gaussianity of the athermal environment.
This also holds for the quasiclassical electronic circuit.
In our model, detailed information of the current fluctuation through the QPC is coded in the statistics of the non-Gaussian noise via ${\mathcal W}({\mathcal Y})$.
The inverse formula to determine ${\mathcal W}({\mathcal Y})$ from ${\mathcal P}_{\rm SS}(\phi)$ can be derived with the aid of Eqs.~(\ref{eq: F_ss}), (\ref{eq: F_classical}), and (\ref{eq: quantum kernel FT}) as
\begin{align}
  {\mathcal W}({\mathcal Y})
  = \frac{1}{\gamma\lambda_{\rm P}}
  \int \frac{d\lambda}{2\pi} e^{-i{\mathcal Y}\lambda} 
  &\left[ 
    \gamma\lambda_{\rm P}
    +\left( {\mathcal T}+\frac{2 k\theta^2}{1+\theta^2\lambda^2}\right)\lambda^2 
    \right. \nonumber \\
    &\left.
    + \lambda\frac{d}{d \lambda} \tilde{{\mathcal F}}_{\rm SS}(\lambda) 
    \right].
  \label{eq: Inverse formula}
\end{align}
The previous result \cite{PhysRevLett.114.090601} is recovered in the classical limit because the parameter $\theta$ in Eq.~(\ref{eq: Inverse formula}) vanishes for $\tau=0$.
The quantum correction is expected to become significant when the temperature of the LC circuit is comparable to the variance of the quantum correction (${\mathcal T} \sim 2k\theta^2$).

\section{Quantum correction and non-Gaussian noise}

\label{sec: results and discussion}

\begin{figure}[b]
  \centering
  \includegraphics[width=\hsize]{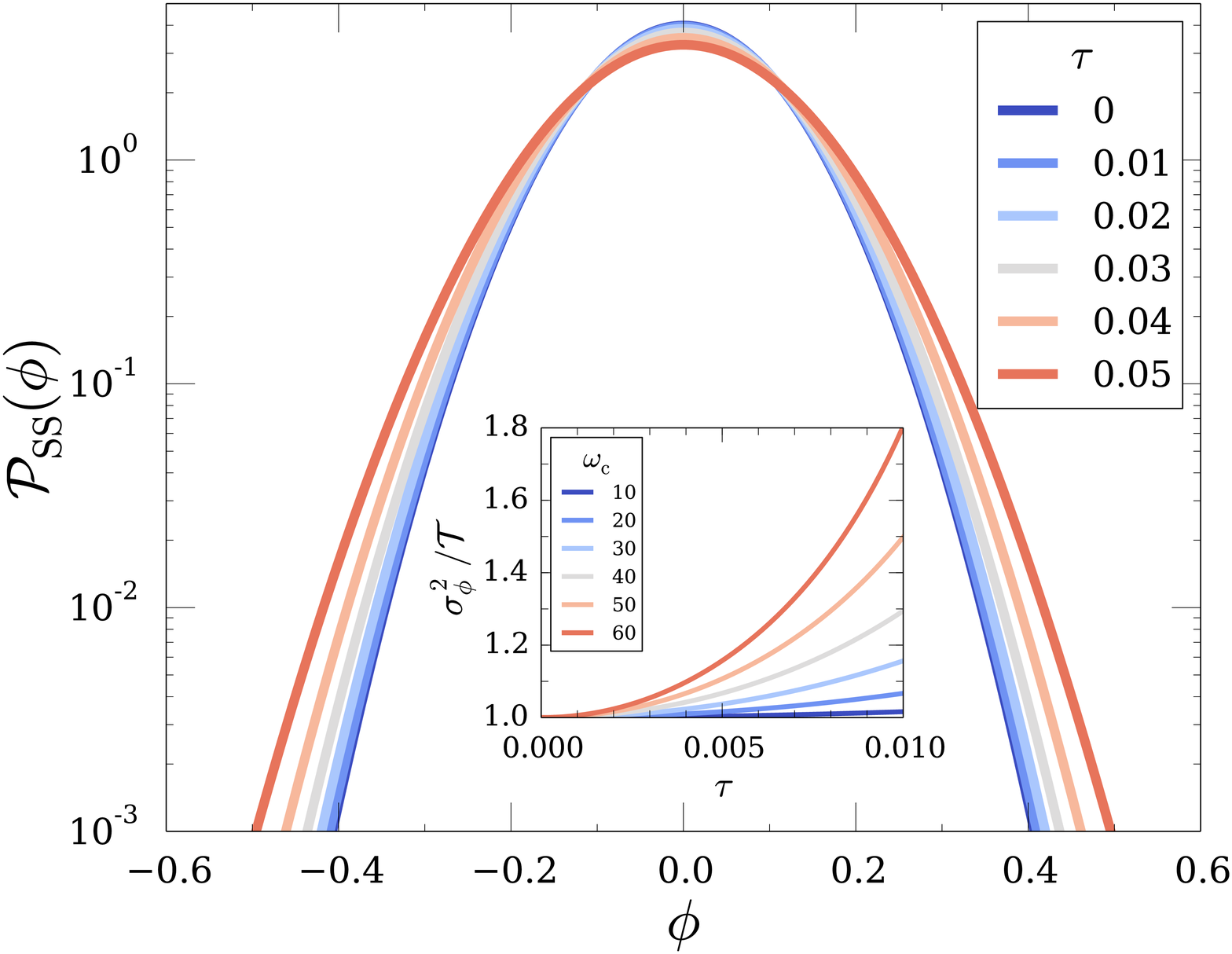}
  \caption{
    (Color online)
    The stationary PDF without the non-Gaussian noise for various values of $\tau$ with
    $\gamma=2.0$, 
    $\mathcal{T}=0.01$,
    and
    $\omega_{\rm c}=10$.
    Inset: the $\tau$-dependence of the variance normalized with its classical value for various values of $\omega_{\rm c}$ with
    $\gamma=2.0$
    and
    $\mathcal{T}=0.01$.
  }
  \label{fig: PDF_tau_w/o_nonGauss}
\end{figure}

We discuss the effect of the quantum correction (\ref{eq: quantum kernel}) in the stationary PDF.
The stationary PDF $\mathcal{P}_{\rm SS}(\phi)$ for the LC circuit without the non-Gaussian noise ($T_{LR}=0$) is shown in Fig. \ref{fig: PDF_tau_w/o_nonGauss}.
In the absence of the non-Gaussian noise, the LC circuit is driven by the thermal and quantum fluctuations.
We examine the effect of the quantum fluctuation for various values of the quantum-mechanical scales $\tau=0$, $0.01$, $0.02$, $0.03$, $0.04$, and $0.05$.
The amplitude of the thermal noise and the cutoff frequency are $\mathcal{T}=0.01$ and $\omega_{\rm c}=10$, respectively.
As is shown in Fig. \ref{fig: PDF_tau_w/o_nonGauss}, the principal effect of the quantum fluctuation is to broaden the stationary PDF.
The behavior can be qualitatively understood
as the increase of the temperature of order $2k\theta^2$.
We note that this interpretation is not perfect because of the $\lambda$-dependence of the quantum correction.
The variance of the PDF $\sigma^{2}_{\phi} \equiv \langle \left( \phi- \langle \phi \rangle \right)^{2} \rangle$ normalized with its classical value $\mathcal{T}$ is plotted in the inset for various values of $\omega_{\rm c}$.
The cutoff modulates the quantum correction by filtering the quantum fluctuation.
The $\tau$ dependence of the variance indicates that
the stationary PDF is rapidly broadened for large cutoff frequencies.
In particular, the variance runs into half of the classical one when the quantum-mechanical scale is about twice the size of the cutoff frequency ($1/\tau \sim 2\omega_{\rm c}$).
This implies that the quantum effect can not be filtered by the cutoff, and the effective temperature becomes about 1.5 times as large as the real temperature.

The non-Gaussian noise produced by the electronic transport through the QPC significantly modifies the stationary PDF.
For simplicity, the transmission coefficient is considered to be so small ($T_{LR} \ll 1$) that the dominant contribution comes from the lowest-order terms ($n=\pm 1$).
In the weak-tunneling regime, the non-Gaussian noise obeys a bidirectional Poisson process \cite{RevModPhys.81.1665}.
The rate parameter $\lambda_{\rm P}$ is reduced to $\lambda_{\rm P}= w_{+1} +w_{-1}$,
and the normalized amplitudes of the non-Gaussian noise take $\pm 1$ with probability $w_{\pm 1}/\lambda_{\rm P}$.
The transition rates $w_{\pm 1}$ can be directly determined by the current and noise measurement (see Appendix \ref{app: transition rate} for the relations).

\begin{figure}[t]
  \centering
  \includegraphics[width=\hsize]{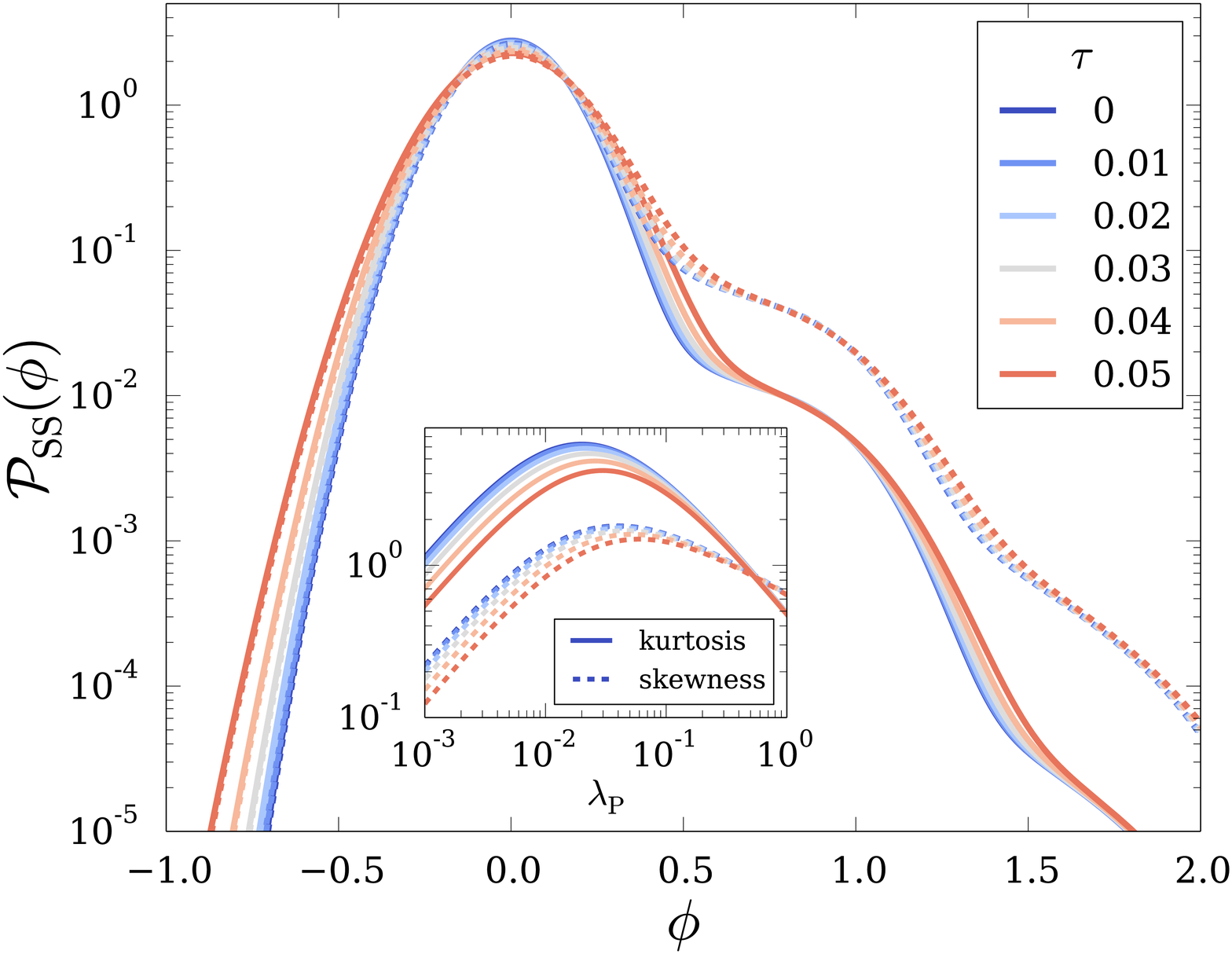}
  \caption{
    (Color online)
    The stationary PDF for the LC circuit coupled to the QPC at large bias voltages.
    The solid (dashed) lines correspond to $T_{LR}=4 \cdot 10^{-6}$ ($T_{LR}=1.6\cdot 10^{-5}$),
    and the other parameters are given by
    $\gamma=2.0$, 
    $\mathcal{T}=0.02$,
    $\omega_{\rm c}=10$, 
    $2\pi\hbar\Omega\beta_{\rm QPC}=10^{3}$,
    and
    $eV/2\pi\hbar\Omega= 10^{3}$.
    Inset: the dependence of the skewness (dashed lines) and the kurtosis (solid lines) on $\lambda_{\rm P}$ for corresponding values of $\tau$.
  }
  \label{fig: PDF_cutoff_nonGaussian_noise}
\end{figure}

With the aid of Eqs.~(\ref{eq: w+}) and (\ref{eq: w-}), the transition probability and the rate parameter at zero-temperature are computed as 
\begin{align}
  \frac{w_{\pm 1}}{\lambda_{\rm P}}&=\theta(\pm eV), 
  \label{eq: W shot noise regime}
\\
  \lambda_{\rm P}&=\frac{T_{LR}}{2\pi} 
  \frac{e|V|}{\hbar\Omega},
  \label{eq: lambda shot noise regime}
\end{align}
respectively.
The sign of the non-Gaussian noise solely depends on that of the bias voltage because electrons flow unidirectionally due to the Fermi statistics.
In the presence of the bias voltage $V$, an attempt rate of an electron-emission event is estimated as $e|V|/2\pi\hbar\Omega$ in the unit time.
The rate parameter is given by the product of the attempt rate and the transmission coefficient $T_{LR}$.
The stationary PDF for the LC circuit coupled to the QPC in the shot noise regime ($2\pi\hbar\Omega\beta_{\rm QPC}=10^{3}$ and $eV/2\pi\hbar\Omega=10^3$) is shown in Fig. \ref{fig: PDF_cutoff_nonGaussian_noise} for various values of $\tau$.
The rate parameter for $T_{LR}=4\cdot 10^{-6}$ ($T_{LR}=1.6\cdot 10^{-5}$) is computed as $\lambda_{\rm P} \simeq 0.004$ ($\lambda_{\rm P} \simeq 0.016$), which is slightly smaller than the variance of the thermal noise (${\mathcal T}=0.02$).
The non-Gaussian noise induces the shift and asymmetry of the PDF; it has a long tail for the positive region of $\phi$.
The characteristic step around $\phi \simeq 1$ originates from the unidirectional single-electron-transfer process which is dominant in the weak-tunneling regime at high-bias voltages.
We note that the precise location of the step is not universal because it depends on the coupling constant $\alpha$ and the friction coefficient $\gamma$ via the normalization of the flux.
However, the presence of the characteristic step clearly shows that the flux of the LC circuit can detect the microscopic electron-transfer events in the QPC.
The quantum correction smoothens the non-Gaussian structures of the PDF.
The behavior is also confirmed by the skewness $\langle \left( \phi- \langle \phi \rangle \right)^{3} \rangle / \sigma^{3}_{\phi}$ and the kurtosis $\langle \left( \phi- \langle \phi \rangle \right)^{4} \rangle / \sigma^{4}_{\phi}-3$), which are plotted in the inset as the dashed and solid lines, respectively.
Nevertheless, the $\tau$-dependence is weak around $\phi \simeq 1$, where the non-Gaussian-noise is dominant over the quantum fluctuation.

\begin{figure}[t]
  \centering
  \includegraphics[width=\hsize]{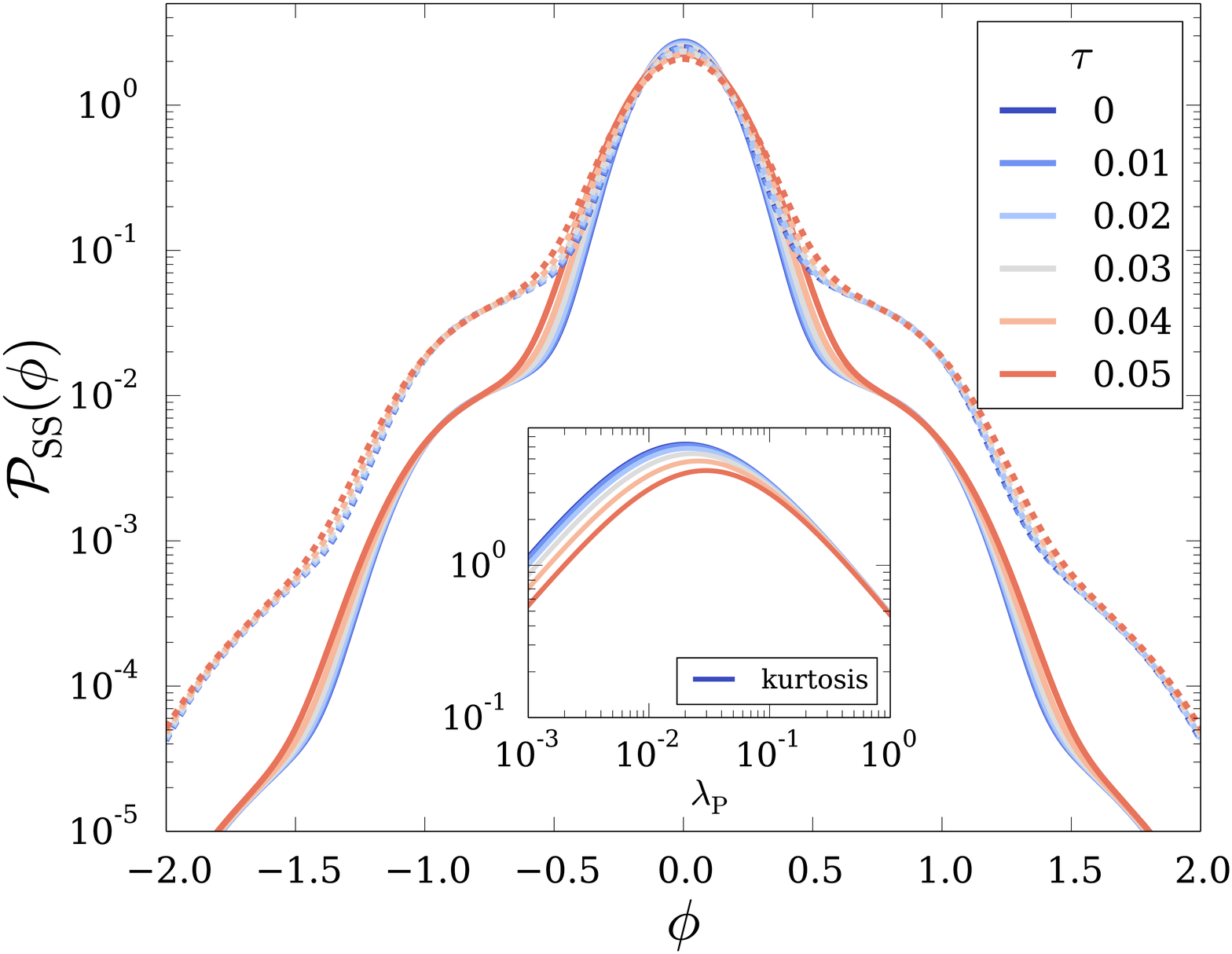}
  \caption{
    (Color online)
    The stationary PDF for the LC circuit coupled to the equilibrium QPC at finite temperatures.
    The solid (dashed) lines correspond to $T_{LR}=4 \cdot 10^{-6}$ ($T_{LR}=1.6\cdot 10^{-5}$).
    Inset: the dependence of the kurtosis on $\lambda_{\rm P}$ for corresponding values of $\tau$.
    Parameters:
    $\gamma=2.0$, 
    $\mathcal{T}=0.02$,
    $\omega_{\rm c}=10$,
    $2\pi\hbar\Omega\beta_{\rm QPC}=10^{-3}$,
    and
    $eV/2\pi\hbar\Omega= 0$.
  }
  \label{fig: PDF_thermal_QPC}
\end{figure}

\begin{figure}[b]
\centering
\includegraphics[width=\hsize]{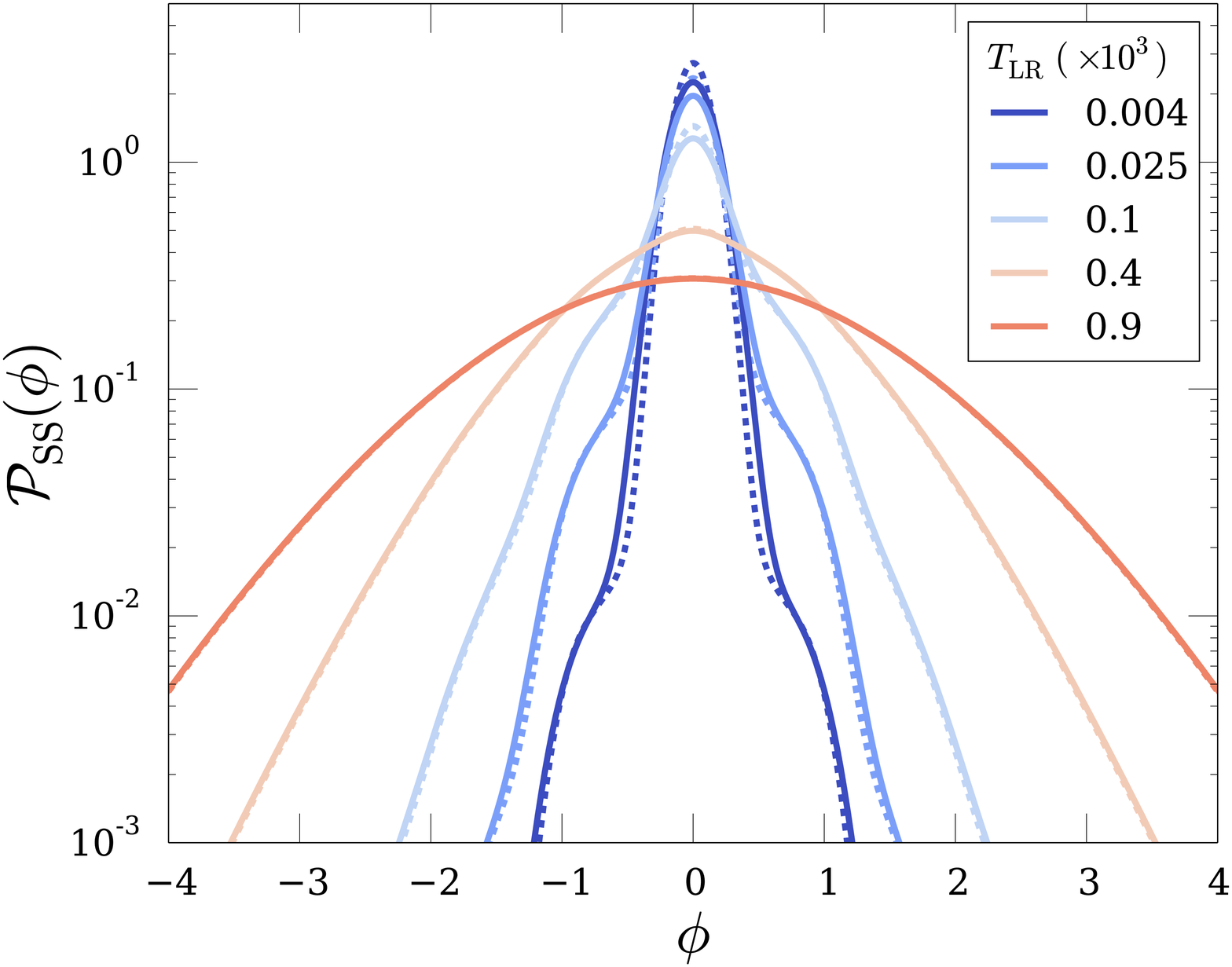}
\caption{
  (Color online)
  The transition of the stationary PDF from the non-Gaussian distribution to the Gaussian one for large values of $T_{LR}$.
  The solid and dashed lines correspond to $\tau=0.05$ and $\tau=0$, respectively.
  The other parameters are the same as in Fig. \ref{fig: PDF_thermal_QPC}.
}
\label{fig: PDF_large_t_LR}
\end{figure}

The QPC at finite temperatures can produce the non-Gaussian noise even without the bias voltage because of its nonlinear coupling to the LC circuit.
The transition probability and the rate parameter are computed as
\begin{align}
  \frac{w_{\pm 1}}{\lambda_{\rm P}}&=\frac{1}{2}, \\
  \lambda_{\rm P}&=\frac{T_{LR}}{\pi\hbar\Omega\beta_{\rm QPC}},
  \label{eq: W finite temperature regime}
\end{align}
respectively.
The non-Gaussian noise with positive and negative amplitudes is generated with the same probability because the current through the equilibrium QPC is bidirectional and unbiased.
The rate parameter is proportional to the product of the temperature and the transmission coefficient.
The stationary PDF for $2\pi\hbar\Omega\beta_{\rm QPC}=10^{-3}$ is shown in Fig. \ref{fig: PDF_thermal_QPC} as the solid lines ($T_{LR}=4 \cdot 10^{-6}$) and the dashed lines ($T_{LR}=1.6 \cdot 10^{-5}$).
The rate parameter is estimated as $\lambda_{\rm P} \simeq 0.008$ and $\lambda_{\rm P} \simeq 0.032$ for each case.
In contrast to the high-bias case shown in Fig.~\ref{fig: PDF_cutoff_nonGaussian_noise}, the stationary PDF exhibits the characteristic structures in both the positive and negative sides around $\phi \simeq \pm1$.
This is because the stationary PDF reflects the forward and backward electron-transfer processes which are allowed in the unbiased QPC.
The non-Gaussian structures are smoothened by the quantum fluctuation as is the case in Fig.~\ref{fig: PDF_cutoff_nonGaussian_noise}.
The reduction of the non-Gaussianity by the quantum fluctuation is also confirmed in the kurtosis plotted in the inset.

The stationary PDF approaches Gaussian as the rate parameter $\lambda_{\rm P}$ is larger:
If the arrival interval of the intermittent noise is much shorter than the decay time, it is piled up to be Gaussian.
The stationary PDF for the relatively large values of the transmission coefficient is shown in Fig. \ref{fig: PDF_large_t_LR}.
The solid lines are the stationary PDF in the presence of the non-Gaussian noise and the quantum fluctuation ($\tau=0.05$), while the dotted lines correspond to the classical limit ($\tau=0$).
The rate parameters are approximately evaluated as $\lambda_{\rm P} \simeq 0.008$, $0.05$, $0.2$, $0.8$, and $1.8$.
The characteristic steps around $\phi\simeq\pm 1$ are smoothened as $\lambda_{\rm P}$ increases.
Moreover, the quantum correction becomes less relevant when the non-Gaussian noise plays a dominant role in determining the PDF.
These behaviors are consistent with the reduction and the $\tau$ independence of the non-Gaussianities, i.e. the skewness and the kurtosis, shown in the insets of Figs.~(\ref{fig: PDF_cutoff_nonGaussian_noise}) and (\ref{fig: PDF_thermal_QPC}) for $\mathcal{T} \ll \lambda_{\rm P}$.

\begin{figure}[b]
\centering
\includegraphics[width=\hsize]{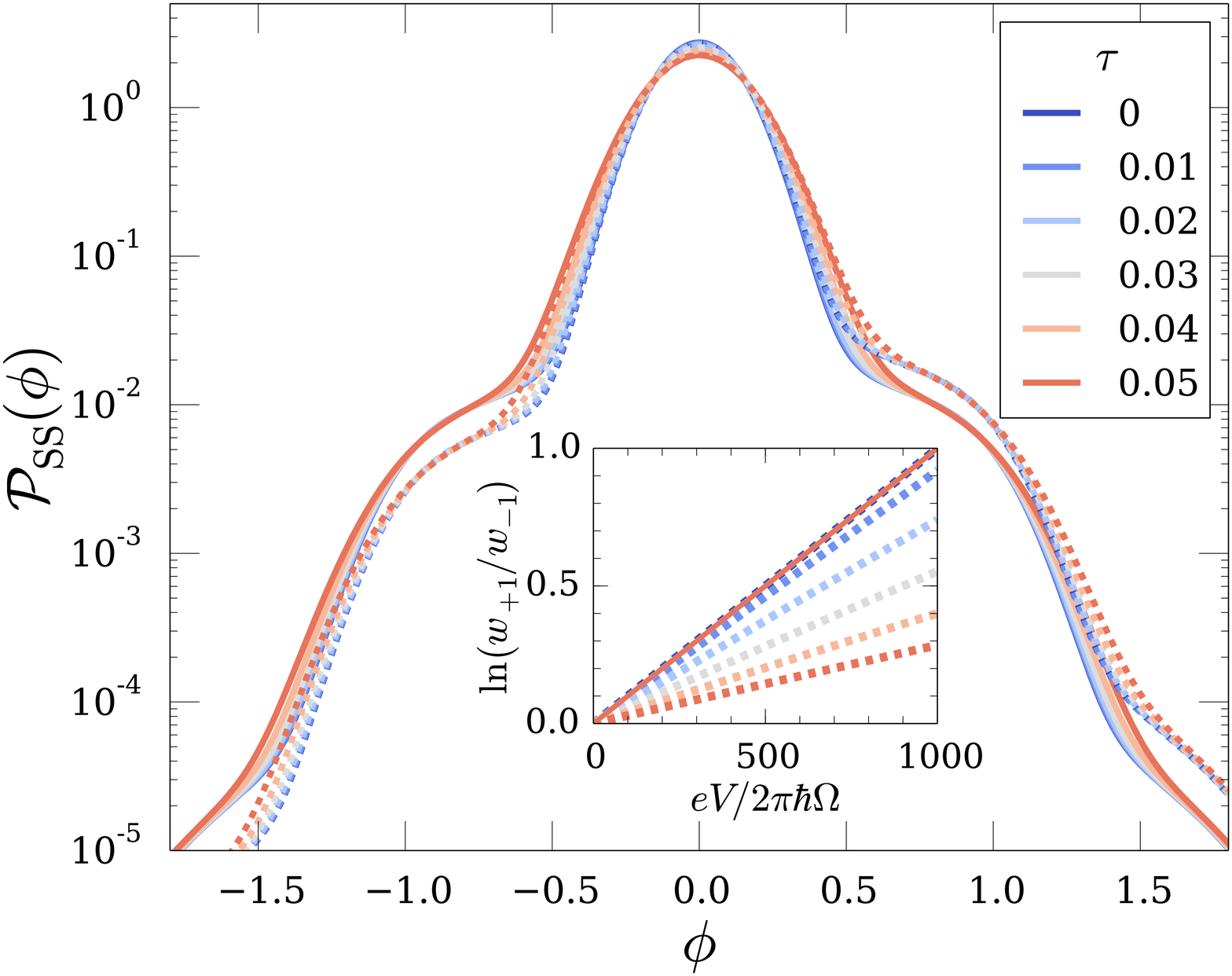}
\caption{
  (Color online)
  The stationary PDF for the LC circuit coupled to the QPC at finite temperature and at bias voltages.
  The solid (dashed) lines correspond to $eV/2\pi\hbar\Omega= 100$ ($1000$).
  Inset: the logarithm of the ratio $w_{+1}/w_{-1}$ estimated by the inverse formula (\ref{eq: Inverse formula}).
  We have obtained the solid curves by properly incorporating the quantum-correction term, which is neglected in the dashed lines.
  Parameters: 
  $\gamma=2.0$, 
  $\mathcal{T}=0.02$,
  $\omega_{\rm c}=10$,
  $T_{LR}=4.0\cdot10^{-6}$,
  and
  $2\pi\hbar\Omega\beta_{\rm QPC}=0.001$.
}
\label{fig: PDF_finite_temperature_voltage}
\end{figure}

Finally, we consider the inverse problem to estimate the transition rates $w_{\pm 1}$ from the stationary PDF ${\mathcal P}_{\rm SS}(\phi)$ by using the inverse formula (\ref{eq: Inverse formula}).
The stationary PDF for the LC circuit coupled to the QPC at finite temperature ($2\pi\hbar\Omega\beta_{\rm QPC}=0.001$) and at finite bias voltages is plotted in Fig. \ref{fig: PDF_finite_temperature_voltage}.
The solid and dashed lines correspond to $eV/2\pi\hbar\Omega= 100$ and $1000$, respectively.
The bias voltage introduces asymmetry of the PDF by modulating the transition rates $w_{\pm 1}$.
Nevertheless, the transition rates must be rigidly related with each other as
\begin{align}
  \label{eq: DB ratio}
  \ln{\left(\frac{w_{+1}}{w_{-1}}\right)}=eV\beta_{\rm QPC},
\end{align}
because of the fluctuation theorem (Appendix \ref{app: transition rate}).
The logarithm of the ratio estimated with the inverse formula (\ref{eq: Inverse formula}) is plotted in the inset of Fig. \ref{fig: PDF_finite_temperature_voltage} as solid lines.
They are on top of each other and perfectly agree with the theoretical prediction (\ref{eq: DB ratio}).
In general, the transition rates which are estimated without the quantum-correction term in Eq.~(\ref{eq: Inverse formula}) are no longer correct.
In order to see how important the quantum-correction term is in the estimation procedure,
we have estimated the ratio by deliberately neglecting the quantum-correction term.
The estimated ratio is plotted as dashed lines in the same inset for various values of $\tau$.
While the curve for the classical limit ($\tau= 0$) is consistent with the fluctuation theorem,
the ratio gradually deviates from the correct value for larger $\tau$.
This result indicates that the quantum correction is important to correctly estimating the transition rate by the inverse formula.

\section{Summary and discussion}
\label{sec: summary}

In this paper, we have studied a dissipative LC circuit inductively coupled to a quantum point contact (QPC).
Under the assumption that the QPC dynamics is much faster than the dynamics of the dissipative circuit,
the system can be considered as a quantum Brownian particle driven by the non-Gaussian noise:
the quantum nature of the dissipative circuit becomes relevant at low temperature,
while instantaneous current through the QPC generates non-Gaussian fluctuation in the circuit.
The quantum fluctuation of the LC circuit at inverse temperature $\beta$ is controlled by the quantum-mechanical scale $\tau=\beta\hbar\Omega/2\pi$ and the Drude cutoff frequency $\omega_{c}=\omega_{D}/\Omega$ with the resonant frequency $\Omega$.
We have evaluated the quantum correction of the steady-state probability density function (PDF) in the quasiclassical regime, where the quantum fluctuation remains week as $\tau<1/\omega_{c} \ll 1$.
The quantum fluctuation is found to smoothen the non-Gaussian structure of the stationary PDF by effectively increasing the temperature.
Nevertheless, the inverse formula, which is useful to infer the non-Gaussian noise from the stationary PDF, can be extended to the classical-quantum crossover regime.
The numerical results indicate that the quantum correction is essential to correctly estimate the statistical property of the non-Gaussian noise with the inverse formula.

The present system is closely related to the detection of the full counting statistics in quantum conductors \cite{levitov1996electron,Nazarov2003}.
The inverse formula obtained in this paper is useful to determine the current distribution from the stationary PDF of the detector circuit in the classical-quantum crossover regime.
It is promising to use the formula for coherent quantum conductors such as Aharonov-Bohm rings \cite{PhysRevLett.104.080602} and superconductors \cite{PhysRevLett.87.197006}.
We note that out formula is crucially based on the assumption that the characteristic time scales of the subsystems are well separated from each other.
It is necessary to go beyond the quasistationary approximation to detect the current fluctuation in the dynamical regime \cite{gabelli2006violation,bocquillon2013coherence}.
In addition, excitations in the electromagnetic environment significantly modify the transport through the mesoscopic conductor in the quantum regime, which is referred to as the dynamical Coulomb blockade phenomena \cite{grabert2013single}.
It is an important future work to fill the gap between our quasiclassical analysis and various quantum approaches to treat such pure quantum cases.
An effective LC circuit whose resonant frequency is of few GHz is experimentally realized with a superconducting device \cite{parlavecchio2015fluctuation}.
The quantum correction studied in this paper comes into play at sub-Kelvin temperatures.

\begin{acknowledgments}
  We thank Kiyoshi Kanazawa, Tomohiko G. Sano, Tatsuhiko Shirai, Takashi Mori, Yasuhiro Utsumi, Kensuke Kobayashi and Seiji Miyashita for useful discussions.
  T.J.S. thanks the warm hospitality of the Yukawa Institute for Theoretical Physics (YITP), Kyoto University where the part of this work is carried out.
  T.J.S. acknowledges financial support provided by the Advanced Leading Graduate Course for Photon Science (ALPS) and YITP.
This work is partially supported by Scientific Grant-in-Aid of JSPS, KAKENHI (No.25287098 and 16H04025).

\end{acknowledgments}

\appendix

\section{Transition rates}
\label{app: transition rate}

The lowest-order terms of the transition rate (\ref{eq: Levitov-Lesovik}) are given by
\begin{align}
  W_{+1}&=\hbar T_{LR} \int \frac{d\omega}{2\pi}
  (1-f_R(\omega)) f_L(\omega) \nonumber \\
  &
  \label{eq: w+}
  =\frac{T_{LR}}{2\pi}\frac{eV}{1-e^{-eV\beta_{\rm QPC}}}, \\
  W_{-1}&=\hbar T_{LR} \int \frac{d\omega}{2\pi}
  f_R(\omega) (1-f_L(\omega)) , \nonumber \\
  \label{eq: w-}
  &=\frac{T_{LR}}{2\pi}\frac{eV}{e^{eV\beta_{\rm QPC}}-1}.
\end{align}
The rate $W_{+1}$ ($W_{-1}$) is proportional to the probability for an electron to be transmitted from the left (right) lead to the right (left) one.

The transition rate $W_{n}$ is directly related to the current fluctuations in the QPC.
According to the full counting statistics, the higher order cumulants of the current fluctuation are generated by introducing the counting field \cite{levitov1996electron} $\chi$ in the hopping amplitude.
If we expand the QPC action in terms of the transmission amplitude and the counting field as
\begin{align}
  i S_{\rm QPC}(\chi)
  &\simeq \left[
    W_{+1}\left( e^{ie\chi }- 1 \right)
    +W_{-1}\left(e^{-ie\chi }- 1 \right) \right] \nonumber \\
  &= \left(W_{+1}-W_{-1}\right)(ie\chi)+\left(W_{+1}+W_{-1}\right)(ie\chi)^2
  \nonumber \\
  & \hspace{20pt} +O(\chi^3),
\end{align}
the current and current noise in the QPC are related to the transition rates $W_{n}$ as
\begin{align}
  I&=\frac{e}{\hbar}\left(W_{+1}-W_{-1}\right), \\
  S&=\frac{e^2}{\hbar}\left(W_{+1}+W_{-1}\right),
\end{align}
respectively.

The forward and backward processes represented by the transition rates $W_{\pm n}$ are mutually related via the fluctuation theorem \cite{RevModPhys.81.1665}.
With the aid of the QPC action (\ref{eq: Levitov-Lesovik}) and the identity
$f_{L}(\omega)(1-f_{R}(\omega))=e^{\beta_{\rm QPC}eV}f_{R}(\omega)(1-f_{L}(\omega))$,
it is straightforward to prove the symmetry of the non-Gaussian characteristic functional
\begin{align}
  \chi_{\rm NG} [\Phi^{\rm q}] 
  = \chi_{\rm NG} \left[-\Phi^{\rm q} + \frac{i \hbar V \beta_{\rm QPC}}{\alpha}\right].
\end{align}
This relation imposes the detailed balance of the transition rates
\begin{align}
  W_{n} = W_{-n}e^{neV\beta_{\rm QPC}},
\end{align}
for $n \in \mathbb{Z}$.

\section{Relation with the Langevin equation}
\label{app: Derivation of Langevin}

The Gaussian noise $\eta_{\rm G}$ can be introduced with the Hubbard-Stratonovich transformation
\begin{align}
  \exp \left[-\varphi^{\rm q}\nu\varphi^{\rm q}\right]
=\int {\cal D}\eta_{\rm G} 
\exp \left[
  -\frac{1}{4} \eta_{\rm G} \nu^{-1} \eta_{\rm G} 
  +i \varphi^{\rm q} \eta_{\rm G} 
\right],
\end{align}
where the statistical property of the auxiliary field $\eta_{\rm G}$ is determined by the noise kernel $\nu$ as
\begin{align}
  \langle \eta_{\rm G}(s) \eta_{\rm G}(s') \rangle = 2\nu(s-s').
\end{align}
If we apply the Hubbard-Stratonovich transformation in the classical limit, the partition function becomes
\begin{align}
  Z=&\int {\cal D}\varphi^{\rm cl}{\cal D}\varphi^{\rm q} {\cal D}\eta_{0} 
  \chi_{\rm NG}[\varphi^{\rm q}] 
 \exp \left[ \int ds  \left(
   -\frac{\gamma}{4 T} \eta^2_{0}(s)
   \right.
   \right. \nonumber \\
 & \left. \left.
   +  i\varphi^{\rm q}(s) 
   \left[
     -\frac{\partial \varphi^{\rm cl}(s)}{\partial s}
     -\frac{\varphi^{\rm cl}(s)}{\gamma}
     +\eta_{0} (s)
   \right]
   \right)
\right],
\end{align}
with the thermal noise $\eta_{0}$.

\section{Derivation of the Master equation}
\label{app:Derivation of Master equation}

We derive the Master equation from the action (\ref{eq: continuous action}) in the classical limit.
The starting point is the discretized action
\begin{align}
  S=\sum_{j} &\left[ -\phi^{\rm q}_{j} \left[\phi_{j} - \phi_{j-1}  
      +\frac{1}{\gamma} \delta_{t} \phi_{j-1}\right] 
    +i\frac{\mathcal{T}}{\gamma} \delta_{t}\left( \phi^{\rm q}_{j} \right)^2 \right. \nonumber \\
    &\left. 
    -i \delta_{t} \lambda_{\rm P}
    \int d{\mathcal Y} \mathcal{W}(\mathcal{Y}) \left( e^{i \mathcal{Y}\phi^{\rm q}_{j}} - 1 \right)  \right].
  \label{eq: discretized action}
\end{align}
The time stride is denoted by $\delta_{t}$,
and the bosonic fields are abbreviated as $\phi^{\rm (q)}_{j} \equiv \phi^{\rm (q)}(j\delta_t)$.
The PDF at the $j$th step is defined in the functional representation as
\begin{align}
  {\mathcal P}_{j} \equiv \Pi^{j-1}_{i'=1}\Pi^{j}_{i''=1} \int {\cal D} \phi_{i'} {\cal D} \phi^{\rm q}_{i''} e^{iS[\phi,\phi^{\rm q}]}.
\end{align}

The PDF ${\mathcal P}_{j}$ is related with ${\mathcal P}_{j-1}$ up to the first order in $\delta_{t}$ as
\begin{align}
  {\mathcal P}_{j} 
  \simeq& \int {\cal D} \delta\phi {\cal D} \phi^{\rm q} 
\biggl[
  1-\frac{i\delta_{t}}{\gamma} \phi^{\rm q} \left(\phi - \delta\phi \right) 
  \nonumber \\
  & \hspace{50pt}
  + \delta_{t} \lambda_{\rm P}
    \int d{\mathcal Y} \mathcal{W}(\mathcal{Y}) \left( e^{i \mathcal{Y}\phi^{\rm q}_{j}} - 1 \right) \biggr] \nonumber \\
&  \times \exp \left[ -
  \left(
    \begin{array}{cc}
      \delta\phi & \phi^{\rm q} 
    \end{array}
  \right)
  \left(
    \begin{array}{cc}
      0 & i/2  \\
      i/2 & \frac{\mathcal{T}}{\gamma} \delta_{t} 
    \end{array}
  \right)
  \left(
    \begin{array}{c}
      \delta\phi \\
      \phi^{\rm q}
    \end{array}
  \right)
 \right]
{\mathcal P}_{j-1},
\label{eq: P_j expansion}
\end{align}
with $\phi\equiv\phi_{j}$, $\delta \phi\equiv\phi_{j}-\phi_{j-1}$, and $\phi^{\rm q}\equiv\phi^{\rm q}_{j}$.
The PDF at $(j-1)$th step ${\mathcal P}_{j-1}$ can be further expanded as
\begin{align}
  {\mathcal P}_{j-1}&=P[\phi-\delta\phi,t_{j}-\delta_{t}], \nonumber \\
  &\simeq\sum^{\infty}_{m=0}  \frac{\left(-\delta\phi\right)^{n}}{n!} \frac{\partial^n {\mathcal P}_{j}}{\partial \phi^n} -\delta_{t} \frac{\partial {\mathcal P}_{j}}{\partial t}.
\end{align}

The functional integration in Eq. (\ref{eq: P_j expansion}) can be performed term by term by using the Wick's theorem with the Gaussian weight
\begin{align}
\exp\left[
  -\left(
    \begin{array}{cc}
      \delta\phi & \phi^{\rm q} 
    \end{array}
  \right)
  \left(
    \begin{array}{cc}
      0 & i/2  \\
      i/2 & \frac{\mathcal{T}}{\gamma} \delta_{t} 
    \end{array}
  \right)
  \left(
    \begin{array}{r}
      \delta\phi \\
      \phi^{\rm q}
    \end{array}
  \right)
\right].
\end{align}
If we denote the expectation value of the quantity ${\mathcal O}$ by $\langle {\mathcal O} \rangle$,
the two-point correlation functions are given by
\begin{align}
  \label{eq: cl-cl}
  &\langle \delta \phi \delta \phi \rangle =\frac{2\mathcal{T}}{\gamma} \delta_{t}, \\
  \label{eq: cross}
  &\langle \delta \phi \phi^{\rm q} \rangle 
  =\langle \phi^{\rm q} \delta \phi \rangle
  =-i, \\
  \label{eq: q-q}
  &\langle \phi^{\rm q} \phi^{\rm q} \rangle =0.
\end{align}
The first line in Eq.~(\ref{eq: P_j expansion}) leads to the Fokker-Planck terms \cite{kamenev2011field}
with the aid of the relations
\begin{align}
  \langle {\mathcal P}_{j-1} \rangle
  &\simeq{\mathcal P}_{j}
  + \frac{1}{2}\langle\delta\phi\delta\phi\rangle \frac{\partial^2 {\mathcal P}_{j}}{\partial\phi^2} 
  -\delta_{t} \frac{\partial {\mathcal P}_{j}}{\partial t} \nonumber \\
  &\simeq{\mathcal P}_{j}
  + \frac{\mathcal{T}\delta_t}{\gamma} \frac{\partial^2 {\mathcal P}_{j}}{\partial\phi^2} 
  -\delta_{t} \frac{\partial {\mathcal P}_{j}}{\partial t} 
  ,
\end{align}
and
\begin{align}
  \langle \phi^{\rm q}
  \left(\phi-\delta\phi \right){\mathcal P}_{j-1}\rangle
  &\simeq - \langle \phi^{\rm q} \delta\phi\rangle 
  \phi 
  \frac{\partial {\mathcal P}_{j}}{\partial \phi} 
  -\langle \phi^{\rm q}\delta\phi\rangle {\mathcal P}_j \nonumber \\
  &\simeq i\frac{\partial\left(\phi {\mathcal P}_{j}\right)}{\partial\phi}.
\end{align}
The non-Gaussian term in Eq.~(\ref{eq: P_j expansion}) can be evaluated as
\begin{align}
  &\delta_{t} \sum^{\infty}_{k=1}\sum^{\infty}_{l=0} 
  \frac{\left( i{\mathcal Y} \right)^k\left(-1\right)^{l}}{k!l!}
  \langle \left( \phi^{\rm q}\right)^k  \left( \phi\right)^l \rangle
  \frac{\partial^l {\mathcal P}_{j}}{\partial \phi^l}
  \nonumber \\
  &\simeq \delta_t \sum^{\infty}_{k=1} \frac{\left( -{\mathcal Y}\right)^k}{k!} \frac{\partial^k {\mathcal P}_{j}}{\partial \phi^k}
  \nonumber \\
  &\simeq \delta_{t} \left( e^{-{\mathcal Y} \frac{\partial}{\partial \phi}} - 1 \right) {\mathcal P}_{j},
\end{align}
up to the leading order in $\delta_{t}$.
We note that the cross correlation is $\mathcal{O}(1)$ while the auto-correlations are zero or higher order terms in $\delta_{t}$ [see Eqs.~(\ref{eq: cl-cl}), (\ref{eq: cross}), and (\ref{eq: q-q})].
This requires the classical components to be contracted with the quantum components of the same number ($k=l$).
The number of the combination of contracting $\phi^{\rm q}$ with $\phi$ is $l!$, which is canceled with the denominator in the first line.

Collecting the Fokker-Planck terms and the non-Gaussian term in the limit $\delta_{t}\rightarrow 0$,
we can derive the Master equation
\begin{align}
  \frac{\partial {\mathcal P}(\phi)}{\partial t} 
  =& \frac{1}{\gamma} \frac{\partial \left( \phi {\mathcal P}(\phi) \right)}{\partial\phi} 
  + \frac{\mathcal{T}}{\gamma} \frac{\partial^2 {\mathcal P}(\phi)}{\partial\phi^2} \nonumber \\
  &+ \lambda_{\rm P}
    \int d{\mathcal Y} \mathcal{W}(\mathcal{Y}) \left( e^{ -\mathcal{Y} \frac{\partial}{\partial \phi}} - 1 \right) {\mathcal P}(\phi).
\end{align}
It is straightforward to generalize these discussions to the multivariate case.


%
%

\bibliography{reference_0402}

\end{document}